\documentclass[preprintnumbers,article,amsmath,amssymb,floatfix,10pt,prd,twocolumn,
superscriptaddress,nofootinbib]{revtex4-2}
\usepackage{bm}
\usepackage{amsfonts}
\usepackage{latexsym}
\usepackage[latin1]{inputenc}
\usepackage{graphicx}
\usepackage{amsmath}
\usepackage{palatino}
\usepackage{mathpazo}
\usepackage{textcomp}
\usepackage{float}
\usepackage{booktabs}
\usepackage{dcolumn}
\usepackage{ragged2e}
\usepackage{hyperref}
\hypersetup{colorlinks,citecolor=blue}
\hypersetup{colorlinks=true,linkcolor=red,filecolor=magenta,    urlcolor=blue}
\usepackage{amsmath}
\usepackage{xcolor}
\usepackage{orcidlink}
\usepackage{epsfig}
\usepackage{caption}
\usepackage{subcaption}
\usepackage{commath}
\def\jnl@style{\it}
\def\aaref@jnl#1{{\jnl@style#1}}

\def\aaref@jnl#1{{\jnl@style#1}}

\def\aj{\aaref@jnl{AJ}}                   
\def\apj{\aaref@jnl{ApJ}}                 
\def\apjl{\aaref@jnl{ApJ}}                
\def\apjs{\aaref@jnl{ApJS}}               
\def\apss{\aaref@jnl{Ap\&SS}}             
\def\aap{\aaref@jnl{A\&A}}                
\def\aapr{\aaref@jnl{A\&A~Rev.}}          
\def\aaps{\aaref@jnl{A\&AS}}              
\def\mnras{\aaref@jnl{Mon.~Not.~Roy.~Astron.~Soc.}}             
\def\prd{\aaref@jnl{Phys.~Rev.~D}}        
\def\prc{\aaref@jnl{Phys.~Rev.~C}}  
\def\prl{\aaref@jnl{Phys.~Rev.~Lett.}}    
\def\qjras{\aaref@jnl{QJRAS}}             
\def\skytel{\aaref@jnl{S\&T}}             
\def\ssr{\aaref@jnl{Space~Sci.~Rev.}}     
\def\zap{\aaref@jnl{ZAp}}                 
\def\nat{\aaref@jnl{Nature}}              
\def\aplett{\aaref@jnl{Astrophys.~Lett.}} 
\def\apspr{\aaref@jnl{Astrophys.~Space~Phys.~Res.}} 
\def\physrep{\aaref@jnl{Phys.~Rep.}}      
\def\physscr{\aaref@jnl{Phys.~Scr}}       
\def\commat{\aaref@jnl{Comm.~Math.~Phys.}}              
\def\science{\aaref@jnl{Science}}               
\def\cqg{\aaref@jnl{Classical Quant.~Grav.}}            
\def\jpcs{\aaref@jnl{JPCS}}                                     
\def\ijmpd{\aaref@jnl{Int.~J.~Mod.~Phys.~D}}                    
\def\grg{\aaref@jnl{Gen.~Relat.~Gravit.}}               
\def\rpp{\aaref@jnl{Rep.~Prog.~Phys.}}          
\def\npa{\aaref@jnl{Nucl.~Phys.~A}}        
\def\lrr{\aaref@jnl{Living Rev.~Rel.}}                   
\def\jcap{\aaref@jnl{J.~Cosmology Astropart.~Phys.}}    
\def\rmp{\aaref@jnl{Rev.~Mod.~Phys.}}   
\def\epjc{\aaref@jnl{Eur.~Phys.~J.~C}} 
\def\plb{\aaref@jnl{~Phy.~Lett.~B}} 
\def\mpla{\aaref@jnl{Mod.~Phy.~Lett.~A}} 
\def\arxiv{\aaref@jnl{arxiv.org}}

\allowdisplaybreaks[1]

\begin{document}
\color{black}       
\title{Reconstruction of $f(Q,T)$ Lagrangian for various cosmological scenario}

\author{Gaurav N. Gadbail\orcidlink{0000-0003-0684-9702}}
\email{gauravgadbail6@gmail.com}
\affiliation{Department of Mathematics, Birla Institute of Technology and
Science-Pilani,\\ Hyderabad Campus, Hyderabad-500078, India.}

\author{Simran Arora\orcidlink{0000-0003-0326-8945}}
\email{dawrasimran27@gmail.com}
\affiliation{Department of Mathematics, Birla Institute of Technology and
Science-Pilani,\\ Hyderabad Campus, Hyderabad-500078, India.}

\author{P.K. Sahoo\orcidlink{0000-0003-2130-8832}}
\email{pksahoo@hyderabad.bits-pilani.ac.in}
\affiliation{Department of Mathematics, Birla Institute of Technology and
Science-Pilani,\\ Hyderabad Campus, Hyderabad-500078, India.}
%
\date{\today}

\begin{abstract}

The variety of theories that can account for the dark energy phenomenon encourages current research to concentrate on a more in-depth examination of the potential impacts of modified gravity on both local and cosmic scales.
We discuss some cosmological reconstruction in $f(Q,T)$ cosmology (where $Q$ is the non-metricity scalar, and $T$ is the trace of the energy-momentum tensor) corresponding to the evolution background in Friedmann-La\^imatre-Robertson-Walker (FLRW) universe. This helps us to determine how any FLRW cosmology can arise from a specific $f(Q,T)$ theory. We use the reconstruction technique to derive explicit forms of $f(Q,T)$ Lagrangian for the different kinds of matter sources and Einstein's static universe. We also formulate the models using several ansatz forms of the $f(Q,T)$ function for $p=\omega \rho$. We demonstrate that several classes of $f(Q,T)$ theories admit the power-law and de-Sitter solutions in some ranges of $\omega$. Additionally, we reconstruct the cosmological model for the scalar field with a specific form of $f(Q,T)$. These new models with cosmological inspiration may impact gravitational phenomena at other cosmological scales.

\end{abstract}

\maketitle


\section{Introduction}

General Relativity has been regarded as the geometry of spacetime since its origin. This view has its foundation in the equivalence principle. Einstein chose to define gravity as a result of the curvature of spacetime. Despite the enormous success of Einstein's General Relativity (GR), cosmic observations introduced new challenges, such as dark energy and dark matter concerns. One of the most important developments in cosmology over the last decade has been the confrontation of observations with the standard $\Lambda$CDM, which led to the discovery of an accelerating universe \cite{Perlmutter/1999,Riess/1998,Riess/2004,Spergel/2007,Koivisto/2006,Daniel/2008}. Despite its accomplishments, this motivated standard model has a severe flaw that stems from the fact that there is a significant discrepancy between the value of $\Lambda$ predicted by any quantum gravity theory and the observational value \cite{Sahni/2000,Padmanabhan/2003,Copeland/2006,Sami/2009}. There are typically two ways to address these issues: modifying the matter sector by including some additional dark components in the universe energy budget and modifying general relativity. In addition to the conventional curvature representation \cite{Buchdahl/1970,Starobinsky/2007}, GR can be associated with torsion or non-metricity. The former corresponds to the Teleparallel Equivalent of GR (TEGR), whereas a Symmetric Teleparallel Equivalent of GR (STEGR) can indeed be supported in flat and torsion-free spacetimes. Their most straightforward modifications are the $f(\mathcal{T})$ theory \cite{Capozziello/2011,Cai/2016} and $f(Q)$ theory \cite{Jimenez/2018,Jimenez/2020,Jimenez/2018a,Harko/2018}. For instance, torsion is considered the field, characterizing gravity in the so-called TEGR. This includes zero curvature and non-metricity, with the Weitzenbock connection as the affine connection. In this context, the virtual objects  are tetrads from which the affine connection, the torsion invariant, and, eventually, the field equations can be derived. While geometry in STEGR has a non-metric connection with both curvature and torsion at zero. Here, the non-metricity $Q$ geometrically explains how a vector's length changes in a parallel transport. Recently, $f(Q)$ gravity has been thoughtfully considered mainly to explain the late-time acceleration and dark energy concerns. For a review, one can check \cite{Atayde/2021}. \\
Furthermore, the $f(Q)$ gravity has been developed to incorporate a non-minimal coupling in the Lagrangian, with the gravitational action generated by an arbitrary function of the non-metricity $Q$ and the trace of energy-momentum tensor $T$ \cite{Xu/2019}. Considerable work has been done in $f(Q,T)$ gravity
using simpler functional forms. Arora et al. demonstrated that $f(Q,T)$ gravity could explain the present cosmic acceleration and provide a feasible solution to the dark energy issue \cite{Arora/2020,Arora/2021,Gadbail/2022}. Additionally,  some quality work based on various cosmological scenarios has been discussed in $f(Q,T)$ gravity theory \cite{Bhattacharjee/2020}.\\
On the other hand, the cosmological reconstruction approach has been designed to precisely recover the properties of $\Lambda$CDM and know the expansion history of the universe through modified theories of gravity. The complexity of field equations, which makes it challenging to acquire exact and numerical solutions that can be compared with observations, hinders studies of the physics of such theories. In the reconstruction technique, it is believed that the expansion history of the universe is understood precisely, and one inverts the field equations to deduce which class of modified theory gives rise to a given flat FRW model. \\
The cosmological reconstruction has been carried out in the framework of $f(R)$ gravity under many scenarios \cite{Nojiri/2006,Capozziello/2006,Nojiri/2009,Goheer/2009a,Dunsby/2010,Carloni/2012} to find realistic cosmology that can represent the evolution of the matter-dominated era to the DE phase. F. Esposito et al. \cite{Esposito/2022} utilize reconstruction methods in $f(Q)$ gravity to examine precise isotropic and anisotropic cosmological solutions. Exact power-law solutions in $f(G)$ gravity are found for a specific class of models as demonstrated by Goheer et al. \cite{Goheer/2009}. Furthermore, similar techniques have been employed to model the cosmic evolution in accordance with the power-law solutions, de Sitter universe, and phantom/non-phantom eras in different extended theories \cite{Elizalde/2010,Sharif/2017,Houndjo/2012,Jamil/2012,Sharif/2014}.\\
We carry out several explicit reconstructions within the framework of $f(Q,T)$ gravity theory, yielding a variety of intriguing results. One noteworthy approach here is to analyze known cosmic history and utilize field equations to construct a specific type of Lagrangian that may reproduce the given evolution background. First, we determine the real values of the Lagrangian $f(Q,T)$ for various types of matter sources and Einstein's static universe. In fact, we discover the appropriate $f(Q,T)$ Lagrangian for modeling the cosmic evolution by the power law and the de-Sitter solutions. We find the gravity models that deviate from $\Lambda$CDM through the ansatz $f(Q,T)= Q+F(T)$, with constant $Q$.\\
This paper is organized as follows. In section \ref{section 2}, we present the general $f(Q,T)$ gravity formalism with the Friedmann equations for the Friedman-Lemaitre-Robertson-Walker universe. We reconstruct the $f(Q,T)$ Lagrangian for different kinds of matter contributions in section \ref{section 3}. By applying the relation $p=\omega \rho$ and the specific ansatz forms of the $f(Q,T)$, we reconstruct $f(Q,T)$ models in section \ref{section 4}. In section \ref{section 5}, we go over the prospect of obtaining gravitational Lagrangians in $f(Q,T)$ that are appropriate for modeling the cosmic evolution implied by the power-law and the de-sitter solutions. In section \ref{section 6}, we reconstruct the cosmological model for the scalar field. Finally, the results are summarised in section \ref{section 7}.

\section{Field equations in $f(Q,T)$ Gravity}
\label{section 2}
The $f(Q,T)$ theory of gravity which introduces an arbitrary function of scalar non-metricity $Q$ and trace $T$ of the energy-momentum tensor, is an intriguing modification to Einstein's theory of gravity. 
The action of $f(Q,T)$ theory coupled with matter Lagrangian $\mathcal{L}_m$ is given by \cite{Xu/2019}
   
\begin{equation}
\label{1}
S=\int \sqrt{-g}\left[\frac{1}{16 \pi }f(Q,T)+\mathcal{L}_m\right] d^4x,
\end{equation}
where $g$ represents the determinant of $g_{\mu \nu}$. The non-metricity and disformation tensor is defined as
\begin{eqnarray}
\label{2}
Q\equiv- g^{\mu\nu}\left(L^\alpha_{\,\,\beta\mu}L^\beta_{\,\,\nu\alpha}-L^\alpha_{\,\,\beta\alpha}L^\beta_{\,\,\mu\nu}\right),\\
L^\lambda_{\,\,\,\,\mu\nu}=-\frac{1}{2}g^{\lambda\gamma}\left(\nabla_{\nu}g_{\mu\gamma}+\nabla_{\mu}g_{\gamma\nu}-\nabla_{\gamma}g_{\mu\nu}\right).
\end{eqnarray}

The non-metricity tensor is defined as the covariant derivative of the metric tensor, and its explicit form is
\begin{equation}
\label{4}
Q_{\alpha\mu\nu}\equiv \nabla_\alpha g_{\mu\nu}.
\end{equation}
with the trace of a non-metricity tensor as
\begin{equation*}
 Q_{\lambda}=Q_{\lambda\,\,\,\,\,\mu}^{\,\,\,\mu}, \quad \quad \tilde{Q}_{\lambda}=Q^{\mu}_{\,\,\,\,\lambda\mu} .
\end{equation*}

The Superpotential $P_{\,\,\mu\nu}^{\lambda}$ is defined as
\begin{equation}
\label{5}
P_{\,\,\,\,\mu\nu}^{\lambda}=-\frac{1}{2}L^\lambda_{\,\,\,\,\mu\nu}+\frac{1}{4}\left(Q^{\lambda}-\tilde{Q^{\lambda}}\right)g_{\mu\nu}-\frac{1}{4}\delta^{\lambda}_{\,\,(\mu\,} Q_{\nu)},
\end{equation} 
giving the relation of scalar nonmetricity as 
\begin{equation}
\label{6}
Q=-Q_{\lambda\mu\nu}P^{\lambda\mu\nu}.
\end{equation}
 The field equations of $f(Q,T)$ theory by varying the action \eqref{1} with respect to the metric tensor is obtained as 
\begin{multline}
\label{7}
-\frac{2}{\sqrt{-g}}\nabla_{\lambda}\left(f_{Q}\sqrt{-g}\,P^{\lambda}_{\,\,\,\,\mu\nu}\right)-\frac{1}{2}f\,g_{\mu\nu}+f_{T}\left(T_{\mu\nu}+\Theta_{\mu\nu}\right)\\
-f_{Q}\left(P_{\mu\lambda\alpha}Q_{\nu}^{\,\,\,\lambda\alpha}-2Q^{\lambda\alpha}_{\,\,\,\,\,\,\,\,\mu}\, P_{\lambda\alpha\nu}\right)=8\pi T_{\mu\nu}.
\end{multline}
The terms used in the above are defined as 
\begin{eqnarray}
\label{8}
\Theta_{\mu\nu} &=& g^{\alpha\beta}\frac{\delta T_{\alpha\beta}}{\delta g^{\mu\nu}}, \quad  T_{\mu\nu}=-\frac{2}{\sqrt{-g}}\frac{\delta(\sqrt{-g}\mathcal{L}_m)}{\delta g^{\mu\nu}},\\
\label{9}
f_T &=& \frac{\partial f(Q,T)}{\partial T}, \quad
f_Q=\frac{\partial f(Q,T)}{\partial Q}.
\end{eqnarray}

We are interested in exploring the cosmological consequences of the $f(Q, T)$ gravitational theory that is defined by a flat Friedmann-Lemaitre-Robertson-Walker (FLRW) metric
\begin{equation}
\label{10}
ds^2=-N^2(t)dt^2+a^2(t)\left( dx^2+dy^2+dz^2\right) ,
\end{equation}
where $N(t)$ is a Lapse function and $a(t)$ is a cosmic scale factor. Since the coincident gauge is fixed during the diffeomorphism, we cannot choose a specific Lapse function. However, the special case of $Q$ theories does allow so because $Q$ retains a residual time reparameterization invariant. Hence, one uses symmetry to set $N(t)=1$ \cite{Jimenez/2018,Jimenez/2020,Jimenez/2018a}. By adopting the coincident gauge, the covariant derivatives reduce to ordinary derivatives. If we adopt a co-moving reference system, with $u^{\mu}=(1,0,0,0)$, then $'= \frac{d}{d\tau}=\frac{d}{dt}$, $\tau$ is a proper time, giving $H=\frac{\dot{a}(t)}{a(t)}$ as the Hubble parameter. For the above metric, the non-metricity scalar is $Q=6H^2$. The trace of energy-momentum tensor is given by $T=-\rho+3p$, where $p$ and $\rho$ are pressure and energy density, respectively.\\
One can also mention that the dynamical evolution of massive particles is not geodesics in $f(Q,T)$ \cite{Xu/2019}, and an additional force does manifest itself as a result of the coupling between $Q$ and $T$. The extra force is orthogonal to matter four-velocity $u^{\mu}$ which is a prerequisite for a physical force. The components that are orthogonal to the four-velocity of the particle can contribute to the equation of motion which is a direct consequence of the co-moving reference frame.

By substituting the above FLRW metric in Eq. \eqref{7} yields the two Friedmann equations of $f(Q,T)$ theory
\begin{eqnarray}
\label{11}
&\frac{f}{2}-6H^2f_Q = 8\pi\rho+f_T(\rho+p),\\
\label{12}
&\frac{f}{2}-2\left[\dot{f_Q} H+f_Q(\dot{H}+3H^2)\right]= -8\pi p.
\end{eqnarray}
Here, dot (.) represent a derivative with respect to cosmic time $t$.\\
We can convert the cosmological evolution equations to a form that resembles standard general relativity, by defining an effective pressure $p_{eff}$ and an effective energy density $\rho_{eff}$ such that 
\begin{equation}
\label{eff1}
 3H^2=8\pi \rho_{eff}=\frac{f}{4f_Q}-\frac{4\pi}{f_Q}\left[\left(1+\frac{f_T}{8\pi}\right)\rho+ \frac{f_T}{8\pi}p \right],  
\end{equation}
\begin{multline}
\label{eff2}
 2\dot{H}+3H^2=-8\pi p_{eff}=\frac{f}{4f_Q}-\frac{2\dot{f}_Q\,H}{f_Q}\\
 +\frac{4\pi}{f_Q}\left[\left(1+\frac{f_T}{8\pi}\right)\rho+\left(2+\frac{f_T}{8\pi}\right)p \right].  
\end{multline}
Hence, follows the effective conservation equation \cite{Xu/2019}
\begin{equation}
\label{15}
\dot{\rho}_{eff}=-3H(\rho_{eff}+p_{eff}).
\end{equation} 
Using equations \eqref{eff1} and \eqref{eff2}, we get the effective equation of state parameter as
\begin{equation}
\label{eff3}
    \omega_{eff}=\frac{p_{eff}}{\rho_{eff}}=-1-\frac{\dot{H}}{H^2}
\end{equation}

 \section{Reconstruction of $f(Q,T)$ cosmology}
 \label{section 3}
We shall now discuss the solutions of \eqref{11} and \eqref{12} pertinent from a cosmological perspective. We aim to demonstrate that every cosmic epoch: dominated by matter, radiation, or dark energy, may be developed  in a model of  $f(Q,T)$ in the upcoming subsections. Adapting the approach of model reconstruction, we consider the functional form of $f(Q, T)$ as 
\begin{equation}
\label{17}
f(Q,T)=f_{1}(Q)+f_{2}(T),
\end{equation}
where $f_{1}$ is a function of $Q$ and $f_{2}$ is a function of $T$.\\

The additive separable model above includes vastly different cosmological limits, such as: STEGR ($f_{1}= Q$ and $f_{2}=0$), $\Lambda$CDM ($f_1+f_2 = 2\Lambda$), $f(Q)$ gravity ($f_2= 0$), STEGR with a modification ($f_1 \neq Q$) allowing the $f_1(Q)$ and $f_2(T)$ function to fully capture the behavior of the effective fluid component. This kind of model has the advantage of yielding a decoupled system of ordinary differential equations for the $f_{1}(Q)$ and $f_{2}(T)$ functions that are simpler to solve.

\subsection{Models for dust $(p=0)$}
Here, the reconstruction for the dust fluid will be discussed. For this case, the trace of energy-momentum tensor (EMT) is $T=-\rho$. 
Using Eq. \eqref{17} in Eq. \eqref{11}, we get
\begin{equation}
\label{18}
\left(\frac{f_1}{2}-Qf_{1Q}\right)=-\left(\frac{f_2}{2}+8\pi T+T f_{2T}\right).
\end{equation}
One can observe that the left-hand side is entirely dependent on $Q$, whereas the right-hand side is a function of $T$. So, due to in-dependency, both sides must be equal to constant (say $\lambda$) using the separation of variables approach. Hence, we have two ordinary differential equations
\begin{eqnarray*}
\frac{f_1}{2}-Q f_{1Q} &=&\lambda, \\
\frac{f_2}{2}+8\pi T+Tf_{2T}&=&-\lambda.
\end{eqnarray*}

Solutions to the above differential equations are
\begin{eqnarray}
\label{19}
f_1(Q) &=&2 \lambda +c \sqrt{Q},\\
\label{20}
f_2(T)&=&-2 \lambda -\frac{16\pi  T}{3}+\frac{c_1}{\sqrt{T}},
\end{eqnarray}
where $c$ and $c_1$ are integrating constants. Substituting in the functional form \eqref{17}, we obtain
\begin{equation}
\label{21}
f(Q,T)=c\sqrt{Q}-\frac{16\pi  T}{3}+\frac{c_1}{\sqrt{T}}.
\end{equation}
One can deduce from the observation that the model in \eqref{21} is a real-valued function of positive non-metricity scalar and the trace of EMT (and positive energy density) if $c_1=0$. It implies the fact that there are classes of real-valued functions $f(Q,T)=c\sqrt{Q}-\frac{16\pi  T}{3}$ other than GR that may simulate a dust-like expansion history.

\subsection{Models for perfect fluid with EoS $p=-\frac{1}{3}\rho$}

In this scenario, we reconstruct the $f(Q,T)$ Lagrangian for an accelerating universe. The EoS value $\omega=-\frac{1}{3}$, which is near the limit of the set of matter fields obeying the strong energy condition, is physically intriguing. The trace of energy-momentum tensor (EMT) here becomes $T=-2\rho$. \\
Again Using Eq. \eqref{17} in Eq. \eqref{11}, we have
\begin{equation}
\label{22}
\left(\frac{f_1}{2}-Qf_{1Q}\right)=-\left(\frac{f_2}{2}+4\pi T+\frac{T}{3} f_{2T}\right)
\end{equation}

The solution of the above differential equation \eqref{22} is 
\begin{eqnarray}
\label{23}
f_1(Q)=2 \mu +c \sqrt{Q},\\
\label{24}
f_2(T)=-2 \mu -\frac{24 \pi  T}{5}+\frac{c_2}{T^{3/2}},
\end{eqnarray}
where $c$ and $c_2$ are integrating constants. Here, $\mu$ is a separation constant.\\
The obtained $f(Q,T)$ model is
\begin{equation}
\label{25}
f(Q,T)=c\sqrt{Q} -\frac{24 \pi  T}{5}+\frac{c_2}{T^{3/2}} .
\end{equation}
One can deduce from the observation that the model in \eqref{25} is a real-valued function of positive non-metricity scalar and the trace of EMT (and positive energy density) if $c_2=0$. The obtained explicit form of the $f(Q,T)=c\sqrt{Q} -\frac{24 \pi  T}{5}$ model is beneficial for studying the acceleration scenario of the universe. 

\subsection{Models for $\Lambda$CDM $(p=-\rho)$}
Let us now consider the standard case for an accelerating universe, i.e., the $\Lambda$CDM. For this, we use a Friedmann equation \eqref{11}, which reduces to
\begin{equation}
\label{26}
\frac{f}{2}-Qf_{Q}=8\pi \rho_{\Lambda}.
\end{equation}

Further, one can rewrite the above equation in terms of density parameters, 
\begin{equation}
\label{27}
\frac{f}{2}-Qf_{Q}=4 \pi Q\,\Omega_{\Lambda}.
\end{equation}
It can be observed that the solution of the differential equation \eqref{27} is purely a function of $Q$.
\begin{equation}
\label{28}
f(Q,T)=-8\pi Q\, \Omega_{\Lambda} +c_3 \sqrt{Q},
\end{equation} 
where $c_{3}$ is an integration constant.
Consequently, the $\Lambda$CDM expansion can be recreated in the modified gravity in the absence of a cosmological constant. In this instance, the $f(Q,T)$ theory is limited to a general relativity equivalent  $f(Q)$ theory.

\subsection{Einstein static universe in $f(Q,T)$ gravity}

The Einstein static universe is described as $H=0$. Hence, by definition of non-metricity, we have $Q=0$. We get the required Friedmann equation \eqref{11} as
\begin{equation}
\label{29}
\frac{f}{2}-8\pi\rho-f_T(\rho+p)=0.
\end{equation}
The dust ($p=0$) results in the trace of EMT being $T=-\rho$, which simplifies the above equation expressed as 
\begin{equation}
\label{30}
\frac{f}{2}+8\pi T+T f_T=0.
\end{equation} 
In general, we obtain the particular solution purely in $T$, 
\begin{equation}
\label{31}
f(Q,T)=-\frac{16 \pi  T}{3}+\frac{c_4}{\sqrt{T}}.
\end{equation}
where $c_4$ is an integration constant. For a real-valued solution, we have $c_4=0$, i.e., $f(Q,T)=-\frac{16 \pi  T}{3}$. 

\section{Reconstruction of $f(Q,T)$ models for EoS $p=\omega \rho$}
\label{section 4}
In this section, we reproduce $f(Q,T)$ models for the perfect fluid satisfying $p=\omega \rho$, where $\omega$ is a constant. The trace of EMT is $T=\rho(-1+3\omega)$. Here, we consider the particular ansatz forms of the $f(Q,T)$ functional forms as follows:  
\begin{itemize}
    \item[A)] $g(Q)+h(T)$, \quad \quad  B) $-Q+h(T)$,\\
    \item[C)] $Q\,h(T)$ ,   \quad \quad \quad \quad  D)\,\, $T\,g(Q)$.
\end{itemize}

Model $B)$ is equivalent to general relativity theory as $f(Q,T)=-Q$ at $h(T)=0$. The coupling possibilities in models $C)$ and $D)$ serve as modification terms to the STEGR Lagrangian. Model $C)$ is fundamentally different since it cannot recover STEGR and is non-trivial. The resultant Lagrangian would therefore characterize cosmological processes without requiring STEGR.  

\subsection{$f(Q,T)=g(Q)+h(T)$}
 We apply additive separable models of $f(Q,T)$ to the Friedmann equation \eqref{11}, which yields a separable partial differential equation as
\begin{equation}
\label{32}
\left(\frac{g}{2}-Q\,g_{Q}\right)=-\left(\frac{h}{2}-8\pi \rho- h_{T}(\rho+p)\right).
\end{equation}
Here, $g_Q$ and $h_T$ represents derivatives with respect to $Q$ and $T$, respectively.\\
Using the relation $\rho=\frac{T}{\left(-1+3\omega \right)}$, the above equation can be written as
\begin{equation}
\label{33}
\left(\frac{g}{2}-Q\,g_{Q}\right)=-\left(\frac{h}{2}-\frac{8\pi}{\left(-1+3\omega \right)} T- h_{T}\frac{1+\omega}{\left(-1+3\omega \right)}T\right)=\lambda.
\end{equation}
where $\lambda$ is a separating constant. The left-hand side and the right-hand side are independent functions of $Q$ and $T$, respectively. Assuming $\lambda \neq 0$, we note that we have a homogeneous differential equation with the following analytical solutions:

\begin{eqnarray}
\label{34}
g(Q) &=&2\lambda+c_5\sqrt{Q},\\
\label{35}
h(T)&=& -\frac{16\pi T}{3-\omega}-2\lambda+c_6\left[2T(1+\omega)\right]^{\frac{-1+3\omega}{2(1+\omega)}}.
\end{eqnarray}
where $c_5$ and $c_6$ are integration constants and can be fixed by using initial conditions on $g$ and $h$.\\
The obtained $f(Q,T)$ model is
\begin{equation}
\label{36}
f(Q,T)=c_5\sqrt{Q}-\frac{16\pi T}{3-\omega}+c_6\left[2T(1+\omega)\right]^{\frac{-1+3\omega}{2(1+\omega)}}.
\end{equation}

\subsection{$f(Q,T)=-Q+h(T)$}

Now, we concentrate on a specific type of $f(Q,T)$ model for which the Lagrangian is expressed as a minimal coupling between the non-metricity scalar and a function of the trace of EMT. It is possible to reduce the Friedmann equation \eqref{11} for this ansatz form of $f(Q,T)$ model.
\begin{equation}
\label{37}
\frac{h(T)}{2}+\frac{Q}{2}=\frac{8\pi T}{\left(-1+3\omega \right)}+\frac{1+\omega}{\left(-1+3\omega \right)}T\,h_T.
\end{equation}
Here, we treat $Q$ as a constant and obtain the solution of the above differential equation. 
\begin{equation}
\label{38}
h(T)=-Q-\frac{16\pi T}{3-\omega}+c_7\left[2T(1+\omega)\right]^{\frac{-1+3\omega}{2(1+\omega)}},
\end{equation}
where $c_7$ is an integration constant.\\
Consequently, the $f(Q,T)$ functional form is given by
\begin{equation}
\label{39}
f(Q,T)=-2Q-\frac{16\pi T}{3-\omega}+c_7\left[2T(1+\omega)\right]^{\frac{-1+3\omega}{2(1+\omega)}}.
\end{equation}

\subsection{$f(Q,T)=Q h(T)$}
The following type of $f(Q,T)$ examines if a function $h(T)$ may satisfy a re-scaling on $Q$. The equation for this paradigm using \eqref{11} is obtained as below.
\begin{equation}
\label{40}
-\frac{Q\,h(T)}{2}=\frac{8\pi T}{\left(-1+3\omega \right)}+\frac{1+\omega}{\left(-1+3\omega \right)}Q\,T\,h_T
\end{equation}
In order to get the solution of equation \eqref{40}, we treat $Q$ as a constant. The analytical solution reads
\begin{equation}
\label{41}
h(T)=-\frac{16\pi T}{Q(1+5\omega)}+c_8\left[2T(1+\omega)\right]^{\frac{1-3\omega}{2(1+\omega)}},
\end{equation}
where $c_8$ is an integration constant. Therefore, the functional form of $f(Q,T)$ becomes\\
\begin{equation}
\label{42}
f(Q,T)=-\frac{16\pi T}{(1+5\omega)}+c_8 Q\left[2T(1+\omega)\right]^{\frac{1-3\omega}{2(1+\omega)}}
\end{equation}

\subsection{$f(Q,T)=T\, g(Q)$}
We now explore the case of an arbitrary function of $Q$ fixing $T$
as a scalar. In this case, we get the desired equation from \eqref{11} as  
\begin{equation}
\label{43}
\frac{T\,g(Q)}{2}-Q\,T\,g_Q=\frac{8\pi T}{\left(-1+3\omega \right)}+\frac{1+\omega}{\left(-1+3\omega \right)}T\,g(Q).
\end{equation}
Here, considering $T\neq 0$ gives 
\begin{equation}
\label{44}
-Q\,g_Q+\frac{\omega-3}{2(-1+3\omega)}g(Q)=\frac{8\pi }{3\omega-1}.
\end{equation}
The general solution of \eqref{44} is obtained as  
\begin{equation}
\label{45}
g(Q)=\frac{16\pi }{\omega-3}+c_9\left[Q(2-6\omega)\right]^{\frac{\omega-3}{2-6\omega}},
\end{equation}
where $c_9$ is an integration constant.\\
The solution reproduces the following $f(Q,T)$ model.
\begin{equation}
\label{46}
f(Q,T)=\frac{16\pi T}{\omega-3}+c_9\,T\left[Q(2-6\omega)\right]^{\frac{\omega-3}{2-6\omega}}
\end{equation}

\subsection{Validity of the above models}

We present two arguments $T(1+\omega)>0$ and $T(1+\omega)<0$ in order to discuss the viability of Models (A), (B), and (C).\\
Here, $T(1+\omega)>0$ gives two conditions:

\begin{itemize}
\justifying
\item $T= \rho \left(-1+3 \omega \right)>0$ and $1+\omega >0$ which gives $\omega >\frac{1}{3}$ and $\omega>-1$, whose intersection gives $\omega >\frac{1}{3}$ which may not be useful in investigating accelerating scenario.
\item $T= \rho \left(-1+3 \omega \right)<0$ and $1+\omega <0$ which gives $\omega <\frac{1}{3}$ and $\omega>-1$, whose intersection gives $\omega <-1$, which is viable to study the accelerated scenario.
\end{itemize} 
Next, $T(1+\omega)<0$ gives two conditions:
\begin{itemize}
\justifying
\item  $T= \rho \left(-1+3 \omega \right)<0$ and $1+\omega >0$ which gives $\omega <\frac{1}{3}$ and $\omega>-1$, whose intersection gives $-1<\omega <\frac{1}{3}$ which can be studied for investigating accelerating scenario only when $\frac{-1+3 \omega}{2(1+\omega)} \in \mathcal{Z}$ .
\item $T= \rho \left(-1+3 \omega \right)>0$ and $1+\omega <0$  gives $\omega >\frac{1}{3}$ and $\omega<-1$, which is not possible at the same time.
\end{itemize} 

Model D), on the other hand, can be studied for cases: $\left(2-6\omega\right)>0$ or $\left(2-6\omega\right)<0$ since $Q>0$. This gives the model viability for either $\omega<\frac{1}{3}$ or $\omega>\frac{1}{3}$ (with $\frac{\omega-3}{2-6\omega} \in \mathcal{Z}$, where $\mathcal{Z}$ is the set of integer numbers), respectively.

\section{Cosmological solutions in $f(Q,T)$ gravity}
\label{section 5}

In this section, we discuss the possibility of acquiring gravitational Lagrangians $f(Q,T)$ that is suitable for simulating the cosmic evolution presented by the power-law and the de-Sitter solutions.

\subsection{Power law solutions}
It would be fascinating to investigate the existence of exact power
solutions in $f(Q,T)$ gravity theory corresponding to various
phases of cosmic evolution. These solutions represent the decelerated and accelerated cosmic eras, distinguished by the scale factor
\begin{equation}
\label{47}
a(t)=a_{0} t^m, \quad H(t)=\frac{m}{t}. \quad (m>0) 
\end{equation}

It is known that the universe is viewed in its decelerated epoch for $0<m<1$. Furthermore, the accelerated phase is experienced for $m>1$. \\

For the corresponding scale factor, the non-metricity scalar takes the form $Q=6m^2t^{-2}$. Additionally, the conservation equation for $p_{eff}=\omega_{eff}\,\, \rho_{eff}$ lead to 
\begin{equation*}
\rho_{eff}(t)=\rho_0t^{-3m(1+\omega_{eff})} .
\end{equation*}
Substituting Eq. \eqref{47} in \eqref{eff3}, we get the effective EoS parameter as
\begin{equation*}
    \omega_{eff}=\frac{(2-3\,m)}{3\,m}
\end{equation*}
Then we may express the effective density $\rho_{eff}$  as a function of $Q$,
\begin{equation}
\label{48}
\rho_{eff}(Q)=\rho_0\left(\frac{Q}{6\,m^2}\right).
\end{equation}
For the sake of simplicity, we suppose that the function can be written in the following form
\begin{equation*}
f(Q,T)=f_1(Q)+f_2(T).
\end{equation*}
In $f(Q, T)$ gravity, only effective thermodynamical quantities satisfy the conservation equation, so we used the effective field equation \eqref{eff1}. Using the above effective quantities listed above, Eq. \eqref{eff1} becomes
\begin{eqnarray}
\label{49}
8\pi \rho_0\left(\frac{Q}{6\,m^2}\right)f_{1Q}-\frac{f_1(Q)}{4}&=&K,\\
\label{50}
\frac{f_2(T)}{4}-4\pi\left[\left(1+\frac{f_{1T}}{8\pi}\right)\rho+\frac{f_{1T}}{8\pi}p\right]&=&K
\end{eqnarray}
Here, $K \neq 0$ is a separating constant. The solution of the differential equation \eqref{49} is
\begin{equation}
    f_1(Q)=-4\,K+\alpha_0\,Q^{\frac{3\,m^2}{16\pi\rho_0}}
\end{equation}
where $\alpha_0$ is an integration constant. 
In Eq. \eqref{50}, real pressure $p$ and energy density $\rho$ are unknown quantities, so we cannot find the analytical solution of differential equation \eqref{50}. We investigate this differential equation to obtain an analytical solution for three possible cases: Dust, perfect fluid, and the $\Lambda$CDM.  
\begin{itemize}

\item For dust universe with $p=0$ and $\rho=-T$, equation \eqref{50} gives
\begin{equation}
\label{52}
f_2(T)=4\,K-\frac{16\pi\,T}{3}+\frac{\alpha_1}{\sqrt{T}},
\end{equation}
where $\alpha_{1}$ is an integration constant.\\
\item For perfect fluid with pressure $p=-\frac{1}{3}\rho$ and $\rho=-T/2$, we get
\begin{equation}
\label{53}
f_2(T)=4\,K-\frac{24\pi\,T}{5}+\frac{\alpha_2}{T^{3/2}},
\end{equation} 
where $\alpha_{2}$ is an integration constant.\\
\item For $\Lambda$CDM case with $p=-\rho$ and $\rho=-T/4$, equation \eqref{50} gives the solution  
\begin{equation}
\label{54}
f_2(T)=4\,K-4\pi\,T,
\end{equation}
\end{itemize}
 Since the energy density in the present universe is positive, $T$ yields negative values for models \eqref{52}, \eqref{53} and \eqref{54}. However, model \eqref{52} and \eqref{53} take a complex value; therefore, it may be compatible with observations when $\alpha_1$ and $\alpha_2$ vanish. The model \eqref{54}, on the other hand, may be consistent with cosmological observations.

\subsection{de Sitter solutions}

The interesting and well-known de-Sitter cosmic evolution effectively describes the expansion of the universe. According to the idea, the universe constantly expands as matter and radiation have a low energy density compared to vacuum (energy density dark energy dominated epoch). The scale factor of this evolutionary model increases exponentially with the constant Hubble parameter $H(t)=H_{0}$ defined as 

\begin{equation*}
a(t)=a_{0} \,e^{H_{0} t}.
\end{equation*}
 For the aforementioned scale factor, the non-metricity scalar is explicitly written as $Q=6H_{0}^2$. \\
Again, using the same functional form $f(Q,T)=f_1(Q)+f_2(T)$, we have the following expressions by using the Friedmann equation \eqref{11}.
\begin{eqnarray*}
\frac{f_1(Q)}{2}-Q_0 f_{1Q} &=& \tilde{K},\\
\frac{f_2(T)}{2}-\frac{8\pi}{-1+3\omega}T-\left(\frac{1+\omega}{-1+3\omega}\right)Tf_{2T} &=& -\tilde{K}.
\end{eqnarray*}
 We can directly solve the above equations for constant $\tilde{K} \neq 0$ to get 
\begin{eqnarray}
f_1(Q) &=& 2\tilde{K}  +\alpha_3 e^{\frac{Q}{2 Q_0}},\\
f_2(T) &=& -\frac{16\pi T}{3-\omega}-2\tilde{K}+\alpha_4\left[2T(1+\omega )\right]^{\frac{-1+3\omega}{2(1+\omega)}}.
\end{eqnarray}
The $f(Q,T)$ model becomes  
\begin{equation}
\label{57}
f(Q,T)=\alpha_3 e^{\frac{Q}{2 Q_0}}-\frac{16\pi T}{3-\omega}+\alpha_4\left[2T(1+\omega)\right]^{\frac{-1+3\omega}{2(1+\omega)}}
\end{equation}
where $\alpha_3$ and $\alpha_4$ are integration constants. The model works for the same conditions as discussed in section \ref{section 4}.

\section{$f(Q,T)$ models for scalar field}
\label{section 6}
The Lagrangian for a scalar field which is minimally coupled to the background reads \cite{Nojiri/2011}
\begin{equation}
L= -\frac{1}{2} \lambda \phi_{,\mu}\phi^{,\mu},
\end{equation}
Here $\lambda$ is a free parameter. The corresponding expression for the stress-energy tensor is
\begin{equation}
T_{\mu \nu}= -\frac{1}{2} \lambda \left(\phi_{,\mu}\phi_{,\mu}-\frac{1}{2}g_{\mu \nu} \phi_{,\alpha}\phi^{,\alpha}\right).
\end{equation}
Now, the expressions for $T$ and $\Theta$ read
\begin{eqnarray*}
T &=& \frac{1}{2} \lambda \phi_{;\alpha}\phi^{;\alpha},\\
\Theta &=& -3 \lambda \phi_{;\alpha}\phi^{;\alpha}.
\end{eqnarray*}

Eliminating the term $\phi_{;\alpha}\phi^{;\alpha}$ from the above two equations, we have $\Theta=-6T$.
Hence, we can write equation \eqref{7} as

\begin{multline}
\label{78}
-\frac{2}{\sqrt{-g}} g^{\mu \nu}\nabla_{\lambda}\left(f_{Q}\sqrt{-g}\,P^{\lambda}_{\,\,\mu\nu}\right)-2 f -5 Tf_{T} \\
 -f_{Q}g^{\mu \nu} \left(P_{\mu\lambda\alpha}Q_{\nu}^{\,\,\,\lambda\alpha}-2Q_{\,\,\,\mu}^{\lambda\alpha}P_{\lambda\alpha\nu}\right)=8\pi G T.
\end{multline}

We choose $f(Q,T)= Q+F(T)$ with a constant $Q$ to get a general solution
\begin{equation}
f(Q,T)=Q - \frac{8\pi G}{7} T + \frac{C}{T^\frac{2}{5}}, \quad T>0.
\end{equation}
where $C$ is an integrating constant.

\section{Conclusion}
\label{section 7}
Currently, one of the most considerable challenges is explaining late time acceleration. The modified theory of gravity models is an intriguing approach to describing such a late-time acceleration without inserting any exotic matter component into the energy budget of the universe.  Xu et al. \cite{Xu/2019} proposed the generalization of the non-metric $f(Q)$ gravity theory by introducing an arbitrary function of the non-metricity scalar $Q$ and the trace of the energy-momentum tensor $T$. They examined the cosmological evolution using three specific classes of the $f(Q,T)$ model. The specific forms considered ignored the presence of specific mixed terms in the $QT$ or its functions. \\
In this work, we presented the reconstruction of different classes of the $f(Q,T)$ model for various cosmological scenarios. We obtained the $f(Q,T)$ Lagrangian for dust fluid, perfect fluid EoS $p=-\frac{1}{3}\rho$, $\Lambda$CDM case and Einstein's static universe using the additive ansatz form of the $f(Q,T)$ Lagrangian. For consistent observational behavior, it is noted that the explicit $f(Q,T)$ forms produced for all the cases in section \ref{section 3} contains the linear term $T$. In the $\Lambda$CDM scenario, the $f(Q,T)$ theory is reduced to the $f(Q)$ theory, while Einstein's static universe is reduced to a strictly $T$ dependent model.

It turns out that the barotropic EoS $p=\omega \rho$ follows the wider class of theories using four different ansatzes $f(Q,T)$ forms as additive separable,  minimal coupling between the non-metricity scalar and a function of $T$, $Q$ re-scaling and $T$ re-scaling models. The additive separable model includes vastly different cosmological models, such as STEGR ($g=Q$ and $h=0$), $\Lambda$CDM ($g+h = 2\Lambda$), $f(Q)$ gravity ($h= 0$), STEGR with a modification ($g \neq Q$) which enables the $g(Q)$ and $h(T)$ functions to accurately represent the behavior of the effective fluid component. The second model is an example of a successful theory of General Relativity that works well when $h=0$. The $Q$ re-scaling is fundamentally different since it cannot recover STEGR and is non-trivial. The resultant Lagrangian would therefore characterize cosmological processes without requiring STEGR. The models in section \ref{section 4} are viable for different ranges of $\omega$ as discussed.

In section \eqref{section 5}, we proposed that possible cosmological trajectories for a general $f(Q,T)$ may correspond to more complex precise solutions that approximate scale like power-law. The expansion history of the universe is believed to have experienced a stage of decelerated power-law expansion followed by late-time acceleration. Therefore, power-law solutions play an essential role in cosmology, as phases are dominated by matter and radiation that eventually connect to an accelerating phase. The existence of the de-Sitter solutions has also been investigated. Finally, we constructed the $f(Q,T)$ Lagrangian for the scalar field using a specific form of $f(Q,T)$.\\
One can conclude that the reconstruction of the viable $f(Q,T)$ forms can be tested through observational data and reproduce different cosmological scenarios (like the background evolution of the universe and late-time acceleration). Even one can study perturbation in future works that can behave differently at the GR limit.

\section*{Data Availability Statement}
There are no new data associated with this article.

\section*{Acknowledgments}

GNG acknowledges University Grants Commission (UGC), New Delhi, India for awarding Junior Research Fellowship (UGC-Ref. No.: 201610122060). SA acknowledges BITS-Pilani, Hyderabad Campus, India for Institute Fellowship. PKS thanks IUCAA, Pune, India for providing support through visiting associateship program. We are very much grateful to the honorable referee and to the editor for the illuminating suggestions that have significantly improved our work in terms of research quality, and presentation.


\begin{thebibliography}{90}
\bibitem{Perlmutter/1999} S. Perlmutter et al.,  Astrophys. J., \textbf{517} 377 (1999).

\bibitem{Riess/1998} A.G. Riess et al., Astron. J., \textbf{116} 1009 (1998).

\bibitem{Riess/2004} A.G. Riess et al., Astophys. J., \textbf{607} 665-687 (2004).

\bibitem{Spergel/2007} D.N. Spergel et al., Astrophys. J Suppl. \textbf{148}, 175 (2003).

\bibitem{Koivisto/2006} T. Koivisto and D.F. Mota, Phys. Rev. D, \textbf{73}, 083502 (2006).

\bibitem{Daniel/2008} S.F. Daniel, Phys. Rev. D, \textbf{77}, 103513 (2008). 


\bibitem{Sahni/2000} V. Sahni and A. A. Starobinsky, Int. J. Mod. Phys. D \textbf{9}, 373 (2000).

\bibitem{Padmanabhan/2003} T. Padmanabhan, Phys. Rep. \textbf{380}, 235 (2003).

\bibitem{Copeland/2006} E. J. Copeland, M. Sami, and S. Tsujikawa, Int. J. Mod. Phys. D \textbf{15}, 1753 (2006).

\bibitem{Sami/2009} M. Sami, Curr. Sci. \textbf{97}, 887 (2009).




\bibitem{Buchdahl/1970} H. A. Buchdahl, Month. Not. R. Astron. Soc. \textbf{150}, 1 (1970).

\bibitem{Starobinsky/2007} A. A. Starobinsky, JETP Letters \textbf{86}, 157 (2007).

\bibitem{Capozziello/2011} S. Capozziello et al., Phys. Rev. D \textbf{84}, 043527 (2011).

\bibitem{Cai/2016} Yi-Fu Cai et al., Rep. Prog. Phys. \textbf{79}, 106901 (2016).

\bibitem{Jimenez/2018} J. Beltran Jimenez et al., Phys. Rev. D \textbf{98}, 044048 (2018).

\bibitem{Jimenez/2020} J.B. Jimenez et al., Phys. Rev. D \textbf{101}, 103507 (2020).

\bibitem{Jimenez/2018a} J. B. Jimenez et al., J. Cosmol. Astropart. Phys \textbf{08}, 039 (2018).

\bibitem{Harko/2018} T. Harko et al., Phys. Rev. D \textbf{98}, 084043 (2018); R. Lazkoz et al., Phys. Rev. D \textbf{100}, 104027 (2019); F. K. Anagnostopoulos, S. Basilakos, and E. N.Saridakis, Phys. Lett. B \textbf{822}, 136634 (2021); N. Frusciante, Phys. Rev. D \textbf{103}, 044021 (2021); G. Gadbail, S. Mandal, and P.K. Sahoo, Phys. Lett. B \textbf{835}, 137509 (2022).

\bibitem{Atayde/2021} L. Atayde, N. Frusciante, Phys. Rev. D \textbf{104}, 064052 (2021); I. Ayuso, R. Lazkoz, and V. Salzano, Phys. Rev. D \textbf{103}, 063505 (2021); R. Lazkoz et al., Phys. Rev. D \textbf{100}, 104027 (2019); D. Zhao, Eur. Phys. J. C \textbf{82}, 1-12, (2022).

\bibitem{Xu/2019} Y. Xu et al., Eur. Phys. J. C \textbf{79}, 708 (2019).

\bibitem{Arora/2020} S. Arora et al., Phys. Dark Univ. \textbf{30}, 100664 (2020).

\bibitem{Arora/2021} S. Arora, A. Parida, and P.K. Sahoo, Eur. Phys. J. C \textbf{81}, 555 (2021).

\bibitem{Gadbail/2022} G. N. Gadbail, S. Arora, and P.K. Sahoo, Phys. Dark Univ. \textbf{37}, 101074 (2022);

\bibitem{Bhattacharjee/2020} S. Bhattacharjee, P.K. Sahoo, Eur. Phys. J. C \textbf{80}, 289 (2020); A. Najera, A. Fajardo, Phys. Dark Univ. \textbf{34}, 100889 (2021); A. Najera, A. Fajardo, J. Cosmol. Astropart. Phys \textbf{03}, 020 (2022); S. Arora et al., J. High Energy Astrophys. \textbf{33}, 1-9 (2022). S. Arora, J.R.L. Santos, P.K. Sahoo Phys. Dark Univ., \textbf{31}, 100790 (2021).
 

\bibitem{Nojiri/2006} S. Nojiri, S.D. Odintsov, Phys. Rev. D \textbf{74}, 086005 (2006)

\bibitem{Capozziello/2006} S. Capozziello, S. Nojiri, S.D. Odintsov, A. Troisi, Phys. Lett. B \textbf{639}, 135 (2006).

\bibitem{Nojiri/2009} S. Nojiri, S.D. Odintsov, D. Saez-Gomez, Phys. Lett. B \textbf{681}, 74 (2009).

\bibitem{Goheer/2009a} N. Goheer, J. Larena, and P. K. S. Dunsby, Phys. Rev. D \textbf{80}, 061301(R) (2009).

\bibitem{Dunsby/2010} P.K.S. Dunsby et al., Phys. Rev. D \textbf{82}, 023519 (2010).

\bibitem{Carloni/2012} S. Carloni, R. Goswami, P.K.S. Dunsby, Class. Quantum Gravity \textbf{29}, 135012 (2012).

\bibitem{Esposito/2022} F. Esposito et al., Phys. Rev. D \textbf{105}, 084061 (2022).

\bibitem{Goheer/2009} N. Goheer et al., Phys. Rev. D \textbf{79}, 121304 (2009).

\bibitem{Elizalde/2010} E. Elizalde et al., Class. Quantum Gravity \textbf{27}, 095007 (2010).

\bibitem{Sharif/2017} M. Sharif, and A. Ikram, Phys. Dark Univ., \textbf{17}, 1-9 (2017).

\bibitem{Houndjo/2012} M.J.S. Houndjo, O.F. Piattella, Int. J. Mod. Phys. D \textbf{21(3)}, 1250024 (2012).

\bibitem{Jamil/2012} M. Jamil et al., Eur. Phys. J. C, \textbf{72}, 1999 (2012).

\bibitem{Sharif/2014} M. Sharif, and M. Zubair, Gen. Relativ. Grav. \textbf{46}, 1723 (2014).

\bibitem{Nojiri/2011} S. Nojiri, and S.D. Odintsov, Phys. Rep. \textbf{505}, 59 (2011).
\end{thebibliography}
\end{document}